\documentclass[aps,prb,twocolumn,superscriptaddress,groupedaddress]{revtex4}
\usepackage{hyperref}
\usepackage{bm}
\usepackage{graphicx}
\usepackage{amsmath}
\usepackage{sidecap}
\usepackage{amsfonts}
\begin{document}
	\title{Majorana fermion arcs and the local density of states of UTe$_2$}
	\author{Yue Yu}
	\affiliation{Department of physics, Stanford University, Stanford, California 94305, USA}
	\author{Vidya Madhavan}
	\affiliation{Department of Physics and Materials Research Laboratory, University of Illinois Urbana-Champaign, Urbana, IL, USA} 
	\author{S. Raghu}
	\affiliation{Department of physics, Stanford University, Stanford, California 94305, USA}
\affiliation{SLAC National Accelerator Laboratory, 2575 Sand Hill Road, Menlo Park, CA 94025, USA}
	\begin{abstract}
		$\text{UTe}_2$ is a leading candidate for chiral p-wave superconductivity, and for hosting exotic Majorana fermion quasiparticles. 
		Motivated by recent STM experiments in this system, we study particle-hole symmetry breaking in chiral p-wave superconductors. We compute the local density of states from Majorana fermion surface states  in the presence of Rashba surface spin-orbit coupling, which is expected to be sizeable in heavy-fermion materials like UTe$_2$. We show that time-reversal and surface reflection symmetry breaking  lead to a natural pairing tendency towards a triplet pair density wave state, which naturally can account for broken particle-hole symmetry. %We compute the resulting local density of states, contributed by the surface-bound states. %Our work points out another exotic behavior of chiral superconductor and is highly relevant for surface probes. We will highlight the relationship between our work and experiments on $\text{UTe}_2$.
	\end{abstract}

	\maketitle
	{\it Introduction - }
	Odd parity superconductors are notable for having a rich pattern of broken symmetries\cite{Leggett1975,Vollhardt2013} and non-trivial topological properties\cite{Qi2011, leijnse2012,Sato2017,Cheung2016}, making them promising avenues for the realization of Majorana fermion modes.  Odd parity superconductivity is inevitably unconventional, arising from electron-electron repulsion rather than from a more conventional electron-phonon pairing mechanism\cite{Sigrist1991}.  Thus, they tend to be realized in strongly correlated systems with complex phase diagrams and competing ordering tendencies. While odd parity superconductivity has been observed in relatively few systems, %while Sr$_2$RuO$_4$ was the leading candidate for odd parity pairing, recent experiments\cite{pustogow2019} have provided striking evidence to the contrary.  
	UTe$_2$ is a strong candidate: it exhibits several striking phenomena including an extraordinarily large upper critical field\cite{aoki2019}, reentrant superconductivity\cite{ran2019} and broken time-reversal symmetry\cite{hayes2020}.  	
	
	Recent scanning tunneling spectroscopy measurements\cite{jiao2020} of UTe$_2$ have revealed an unusual signature that is yet to be explained. STM spectra obtained near step edges in the superconducting phase show broken particle-hole  symmetry (PHS).  This observation is sharply distinct from classical BCS superconductors, which exhibit an emergent PHS at energies below the gap: the difference between adding and removing an electron is a Cooper pair, which is unobservable in a conventional Cooper pair condensate.  Since STM directly measures local density of states, these unusual edge modes may be signatures of putative chiral Majorana modes, and thus, it is of central interest to understand these experimental observations.  

A clue to understanding the question of broken PHS in a superconductor comes from studies of materials with non-zero Cooper pair momentum\cite{FF1964,LO1965,Agterberg2020} which are typically associated with a pair density wave (PDW) modulation. It is well-known that when superconductivity is accompanied by charge modulation, PHS breaking  usually occurs.
It is thus possible that the observations of broken PHS in UTe$_2$  can be explained by the development of inhomogeneous superconductivity near the surface.	In this letter, we show how PHS breaking can emerge from finite momentum Cooper pairing near the surface of an odd parity chiral superconductor. % Essentially, just as Zeeman fields can give rise to finite momentum pairing in {\it even} parity superconductors, the breaking of reflection symmetry can lead to a similar fate for odd-parity states. Since the surface of material always breaks reflection symmetry, the role of a Zeeman field in even parity superconductors is taken up by a symmetry-allowed Rashba spin-orbit coupling (SOC) at the surface. %We present our analysis of the recent STM experiment in one such heavy-fermion superconductor, UTe$_2$. 
%Our calculations show that in the presence of Rashba SOC, a pair density wave (PDW) state is induced at the surface of a chiral superconductor. This inhomogeneous superconducting state breaks PHS and may account for the STM observations. 
	
	{\it Symmetry considerations - }
Since the PDW state is not a generic weak-coupling instability of a Fermi liquid, one cannot predict unambiguously when it might occur.  Nevertheless, there are some reliable rules-of-thumb. When the normal state has both time-reversal and inversion symmetry, the spectrum consists of Kramers degenerate pairs at momenta $\left( \pm \bf k \right)$ and uniform superconductivity is overwhelmingly preferred in the weak-coupling limit. In this case, the BCS superconductor is labeled by a ``pseudospin" degree of freedom stemming from the normal state Kramers' degeneracy. Thus, to tilt the balance in favor of the  PDW state,  either inversion or time-reversal symmetry ought to be broken in the normal state. For instance, a Zeeman field can help stabilize a PDW state in pseudospin singlet superconductors, as is believed to be the case in CeCoIn$_5$\cite{martin2005,Kenzelmann2008}.  Similarly, a PDW state in an odd parity superconductor can be stabilized by breaking inversion (or reflection) symmetry, as we explain below.  

	In the case of UTe$_2$, both time-reversal and inversion symmetries occur in the normal state in the {\it bulk}. However, spatial reflection symmetry is broken at the {\it surface}. This leads to a  Rashba spin-orbit coupling (SOC) which decays in strength away from the surface. If this decay length is sufficiently long compared to the superconducting correlation length, or if the SOC itself is sufficiently strong (as is likely the case in UTe$_2$ due to the large bulk atomic SOC scales), Majorana surface modes will be strongly affected by the reflection symmetry breaking at the surface, which results in a PDW component to the condensate. The symmetry considerations therefore suggest that the STM observations of broken PH in UTe$_2$ may be attributed to non-uniform superconductivity induced at the surface by sizeable Rashba SOC.
	
	{\it Model- }%\label{S2}
	We first provide an explicit example that illustrates how non-uniform superconductivity is induced by local inversion symmetry breaking.  Since the electronic structure of UTe$_2$ remains poorly understood we instead consider a single band effective description.  Let us suppose that the overwhelming pairing tendency is in the odd parity pseudopsin triplet channel without a PDW component.  Deep in the bulk, the superconductor is described by mean-field Hamiltonian  of the following form:
	\begin{eqnarray}
		H &=& H_0+H_{\Delta}+H_{Rashba}   \nonumber \\
		H_0 &=& \sum_{\bf k, \sigma} \left( E_{\bf k   \sigma}  - \mu \right) c^{\dagger}_{\bf k  \sigma} c_{\bf k  \sigma} \nonumber \\
		H_{\Delta}&=& \sum_{\bf k, \sigma, \sigma'} \Delta_{\sigma \sigma'\bf k } c_{ \bf k, \sigma } c_{- \bf k ,\sigma' } + {\rm h.c.}.
	\end{eqnarray}
	We assume that the system has lattice translation symmetry.  Thus, $c_{\bf k  \sigma}$ destroys an electron with crystal momentum $\bf k$, and pseudospin $\sigma$.  The band energies $E_{\bf k \sigma}$ are time-reversal symmetric and appropriate for an orthorhombic crystal, such as that of UTe$_2$.  
	Pseudospin-triplet superconducting states are then characterized by a matrix-valued order parameter $\Delta_{\sigma \sigma'\bf k}$.
	%\begin{equation}
	%	\label{dvector}
	%	\Delta_{\sigma \sigma'\bf k \alpha} = i\left( \sigma^y \vec \sigma  \cdot \vec d_{\alpha}(\bf k) \right)_{\sigma \sigma'}.
	%\end{equation}
	%In what follows, we will ignore the band index and treat an effective single band problem. 

	%In an orthorhombic crystal, superconductivity with broken time-reversal symmetry ordinary requires two transitions since all symmetry allowed gap functions belong to a one-dimensional irreducible representation of the crystalline point group.  In this case, there is first a transition from a normal state to a superconducting state at temperature T$_{c1}$, and then a second transition to a superconductor with broken time-reversal at a lower temperature T$_{c2} <$ T$_{c1}$.  Indeed, precisely such a scenario has been reported in recent experiments\cite{hayes2020}. Let us, therefore, assume that for  $T \ll T_{c2}$, the bulk remains deep within an odd parity superconducting state with broken time-reversal symmetry, which enables us to ignore other competing ground states from the standpoint of the bulk.  
	
	Near the surface, superconductivity is subject to a sizeable Rashba spin-orbit coupling due to the breaking of reflection symmetry. Letting $\hat n$ be the vector normal to a surface, the Rashba SOC Hamiltonian usually takes the form
	\begin{equation}
		H=\lambda \sum_{\bm k \sigma} \hat n \cdot \left( \bm k \times \bm \sigma \right).
	\end{equation}
	In an orthorhombic crystal, such as UTe$_2$, the lack of a fourfold rotational axis results in a less restrictive form of Rashba SOC.  
	Letting $\hat n = \hat y$, the surface Rashba SOC of an orthorhombic system has the form
	\begin{equation}
		H_{Rashba} = -\lambda{}k_xS_z+\lambda'k_zS_x,
	\end{equation}
	with generically distinct values of $\lambda$ and $\lambda'$. In what follows, we will neglect spatial decay of $\lambda, \lambda'$ into the bulk and treat them as constant parameters of the surface Hamiltonian.  %  the spatial depeassume that the spatial dependence of Rashba SOC is unimportant for the chiral surface modes, and take $\lambda$ and $\lambda'$ to be constants. 
	In addressing the effect of these surface Fermi surface distortions on superconductivity, we assume that the superconducting gap scale is small compared to the Rashba coupling, as is usually the case for BCS superconductors.
	
	%Near the surface, however, the situation can be different.  Even for temperatures $T \ll $T$_{c2}$, surface superconductivity is subject to a sizeable Rashba spin-orbit coupling due to the breaking of reflection symmetry at a surface.  Letting $\hat n$ be the unit normal vector associated with the surface, the Rashba coupling is usually taken to be a modification to the kinetic energy of the form 
	%\begin{equation}
	%	H_{Rashba} = \lambda \sum_{\bf k,  \sigma \sigma'} c^{\dagger}_{\bf k,  \sigma} \left[ \hat n \cdot \left( \bf k \times \bf \sigma \right)_{\sigma \sigma'} \right] c_{\bf k,  \sigma'}.
	%\end{equation} 
	%While the Rashba coupling only requires a breaking of spin-rotation and reflection symmetries, it is reasonable to suppose that it would be enhanced near the surface of a  heavy-fermion material with sizeable atomic spin-orbit coupling in the bulk.  
	%In an orthorhombic system, such as UTe$_2$, the Rashba coupling is allowed to take a less symmetric form due to the absence of fourfold rotation symmetry about $\hat n$: for instance for a surface with $\hat n = \hat y$, it has the form
	%\begin{equation}
	%	H_{Rashba} = -\lambda{}k_xS_z+\lambda'k_zS_x,
	%\end{equation}
	%with generically distinct values of $\lambda$ and $\lambda'$.  
	
	\begin{figure}[h]
		\centering		\includegraphics[width=8cm]{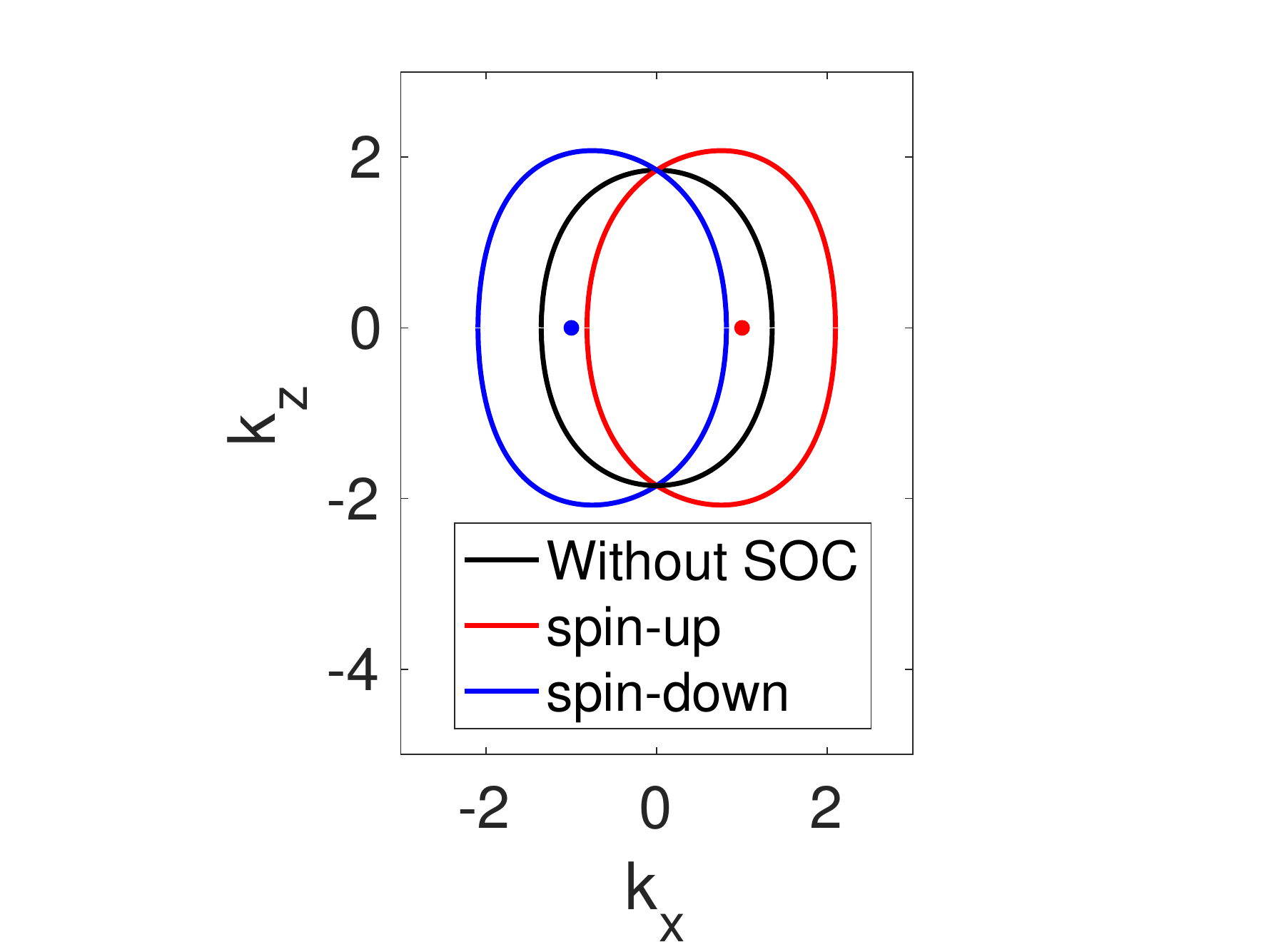}
		\caption{Projected Fermi surfaces onto $(k_x,k_z)$ plane for (Black line)Fermi surface in the absence of Rashba SOC, as well as (Red,Blue) pseudospin-up,down Fermi surface under Rashba SOC. Band dispersion is $E_{\bf k \sigma}=\sum_i\frac{k_i^2}{2m_i}+\sum_i\frac{k_i^4}{2n_i}$ for the plots, with $m_x=1$, $m_y=2$, $m_z=3$, $n_x=20$,  $n_y=40$, $n_z=60$, $\mu=1$, $\lambda=0.8$ and $\lambda'=0$. Blue and red dots are the approximate position for $\bf{q}_{\pm}$. }
		\label{FS0}
	\end{figure}

	The existence of an induced PDW is most clearly demonstrated in the extreme orthorhombic limit where %Let us start with the limit of
	$\lambda'\ll\lambda$.  In this limit, the pseudospin $S_z$ is well approximated as a good quantum number.   %pseudospins correspond to $S_z$ eigenvectors. 
	Moreover, $\Delta_{\uparrow \uparrow}$ and $\Delta_{\downarrow \downarrow}$ condensate now ``live" on separate Fermi surfaces as shown in Fig. \ref{FS0}, and therefore prefer non-zero centers of mass momenta $\bf{q}_{\pm}\equiv(q_{x\pm},0,0)$. And it remains energetically more favorable for the condensate in the $\Delta_{\uparrow \downarrow}$ channel to have zero  center of mass momentum. Since chiral p-wave state breaks TRS, in general $\bf{q}_+\neq-\bf{q}_-$.
	If all three channels are ordered, there will be three distinct condensate momenta, while two condensate momenta are already sufficient for the PDW\cite{LO1965,Agterberg2020}. The difference between those two momenta determines the orientation and periodicity of PDW:
	\begin{equation} \Delta(x)\sim\cos(\frac{1}{2}\Delta\bf{q\cdot{x}})
	\end{equation}
	Since the three pairing channels have different condensate momenta, they will naturally compete with each other. In this $\lambda' \ll\lambda$ limit, if $\Delta_{\uparrow \downarrow}$ channel is suppressed by competition, the remaining competition between $\Delta_{\uparrow \uparrow}$ and $\Delta_{\downarrow \downarrow}$ channel will be weak; therefore it is energetically easier for these two condensates to coexist. Other coexisting scenarios are more difficult to happen, but still possible under certain pairing interactions.
	
	The complementary extreme limit $\lambda'>>\lambda$, can be handled in the same way upon exchanging the $x$ and $z$ coordinates. In terms of the original coordinate system, the resulting PDW state would now have its center of mass momenta along the z-direction. 
	
	As we deviate from the limit $\lambda' \ll \lambda$, the first corrections from $\lambda'$ would be to alter the shape of Fermi surfaces and the momenta of the PDW phase, but we expect the PDW phases to survive to a finite range of $\lambda'/\lambda$.  Similarly, we expect the PDW in the complementary limit to survive up to a finite range of $\lambda/\lambda'$.  %The above two extreme cases naturally extend to two phases with a finite range of $\lambda'/\lambda$.
	Since the two PDW phases in either extreme have distinct center of mass momenta, there are a variety of possibilities for intermediate $\lambda'/\lambda$.  There could be a direct first-order transition where the PDW momenta jump abruptly from one phase to the other, or a coexistence phase with both sets of PDW momenta present.  There could also be an intermediate phase without PDW order.  % there could be at least one intervening phase either with uniform pairing or coexistence between the two PDW phases.  there could be a direct transition or intermediate phase(s) between these two phases. The intermediate phase could be the uniform $q=0$ pairing state, where PDW is lost and PH symmetry is preserved. Otherwise, the intermediate phase could be the bidirectional PDW, where the condensate has a non-zero center of mass momenta in both x and z-direction. 
	The correct scenario needs to be determined by the detail of the Fermi surface and the pairing interactions, while the $\lambda/\lambda'$ ratio is determined by the orientation of the measured surface.  But we can state with certainty that the PDW phases obtained in the limit of extreme orthorhombicity do not require fine-tuning. 
	
	{\it Chiral edge modes and LDOS - }%\label{S6}
	%In this work, we will 
	Next we consider the quasiparticle spectrum in the 
	strongly orthorhombic limit with $\lambda\neq0$ and $\lambda'=0$, assuming all three channels are ordered. Since translational symmetry is broken in the PDW, the Bogoliubov–de Gennes (BdG) Hamiltonian is not block-diagonal, and truncation is needed for concrete computations. To illustrate the essential idea at  a qualitative level, we  truncate the Hamiltonian to the following $4\times4$ form to include just two PDW momenta $q_{+}, q_{-}$: 
	\begin{equation}
		\begin{split}
			&H=\sum_{\bf k}\Phi_{\bf k}^\dagger{}h_{\bf k}\Phi_{\bf k}\\
			&\Phi_{\bf k}^\dagger=[c_{\bf k\uparrow}^\dagger\;\; c_{\bf k+q_-\downarrow}^\dagger\;\;
			c_{\bf -k+q_+\uparrow}\;\;
			c_{\bf -k\downarrow}]\\
			&h_{\bf k}=\left[
			\begin{array}{cccc} \varepsilon_{\bf k\uparrow} & 0 & \Delta_{\uparrow\uparrow}({\bf k}) & \Delta_{\uparrow\downarrow}({\bf k})\\
				0 & \varepsilon_{\bf k+q_-\downarrow} & 0 & \Delta_{\downarrow\downarrow}({\bf k+q_-})\\
				\Delta^*_{\uparrow\uparrow}({\bf k}) & 0& -\varepsilon_{\bf -k+q_+\uparrow}  & 0 \\
				\Delta^*_{\uparrow\downarrow}({\bf k}) &
				\Delta^*_{\downarrow\downarrow}({\bf k+q_-}) & 0 & -\varepsilon_{\bf -k\downarrow}
			\end{array}
			\right]
		\end{split}
	\end{equation}
	with the following simple normal state dispersion: 
	\begin{equation}
		\begin{split}
			&\varepsilon_{\bf k\uparrow}=\frac{k^2}{2m}-\mu-\lambda{}k_x;\;\;\varepsilon_{\bf k\downarrow}=\frac{k^2}{2m}-\mu+\lambda{}k_x
		\end{split}
	\end{equation}	
	The anti-commutation relationship imposes the following constraint on the pairing function:
	\begin{equation}
	\begin{split}
	&\Delta_{\uparrow\uparrow}({\bf k})=-\Delta_{\uparrow\uparrow}({\bf q_+-k})\\
	&\Delta_{\uparrow\downarrow}({\bf k})=-\Delta_{\uparrow\downarrow}({\bf -k})\\
	&\Delta_{\downarrow\downarrow}({\bf k})=-\Delta_{\downarrow\downarrow}({\bf q_--k})
	\end{split}
	\label{e2}
	\end{equation}
	It should be noted that much larger matrices and a realistic band structure are required for future quantitative analysis. The result for $8\times8$ truncation can be found in the appendix.
	
	Using the above Hamiltonian, we compute the surface bound state spectrum and the local density of states to validate the qualitative picture above of broken PHS. We will compare the results in (1) the uniform $\bf q=0$ state without Rashba SOC, and (2) the non-uniform PDW state with Rashba SOC. For simplicity, we take the same ``$p_x+ip_y$" pairing state for both cases:
	\begin{equation}
		\begin{split}
			&\Delta_{\uparrow\uparrow}({\bf k})=\Delta_1k_z\\
			&\Delta_{\uparrow\downarrow}({\bf k})=\Delta_{\downarrow\uparrow}({\bf k})=\Delta_2k_x+i\Delta_3k_y\\
			&\Delta_{\downarrow\downarrow}({\bf k})=\Delta_4k_z,
		\end{split}
	\end{equation}
	with real coefficient $\Delta_i$. 
	
	\begin{figure}[h]
		\centering
		\includegraphics[width=6cm]{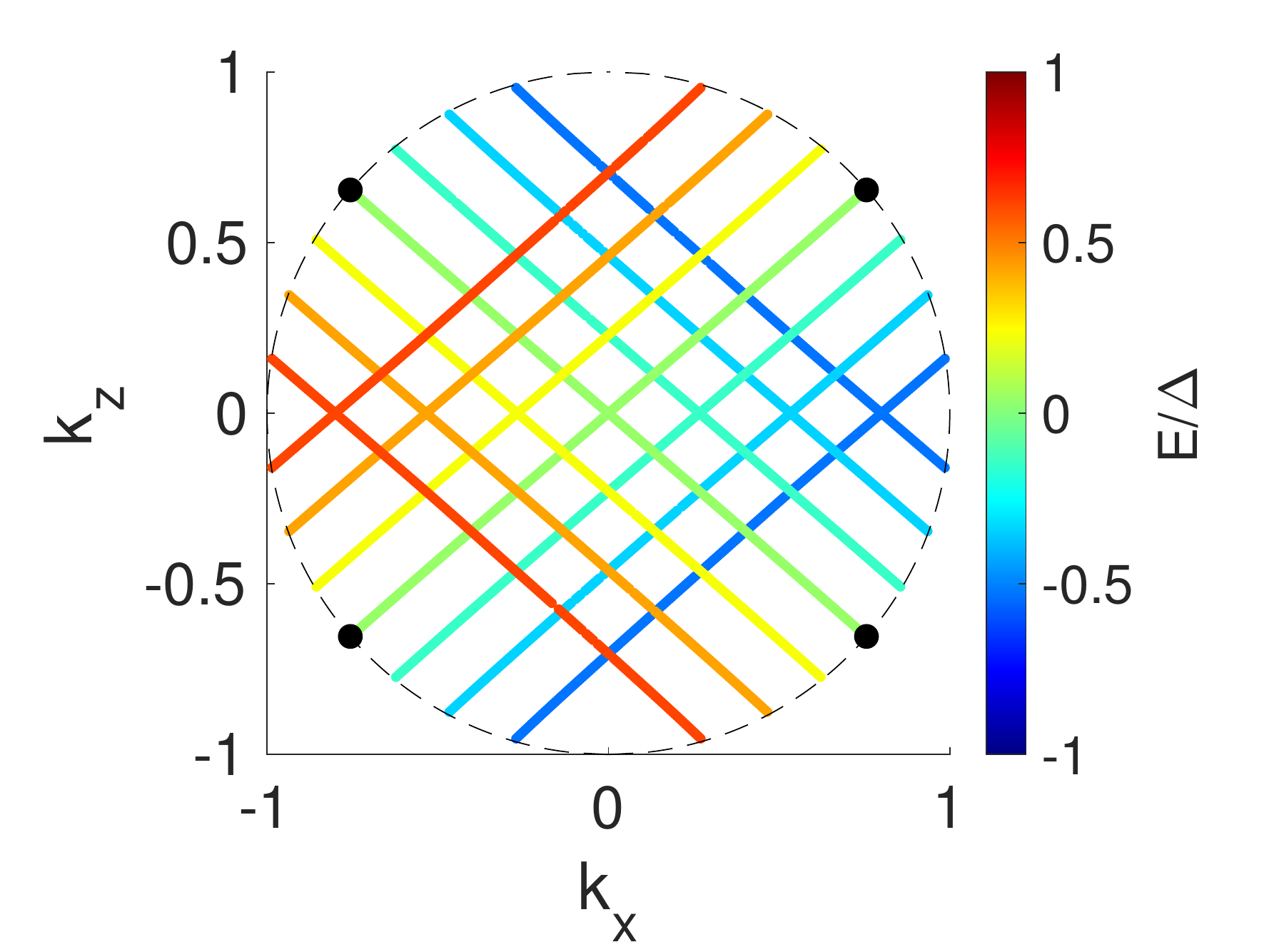}
		\includegraphics[width=6cm]{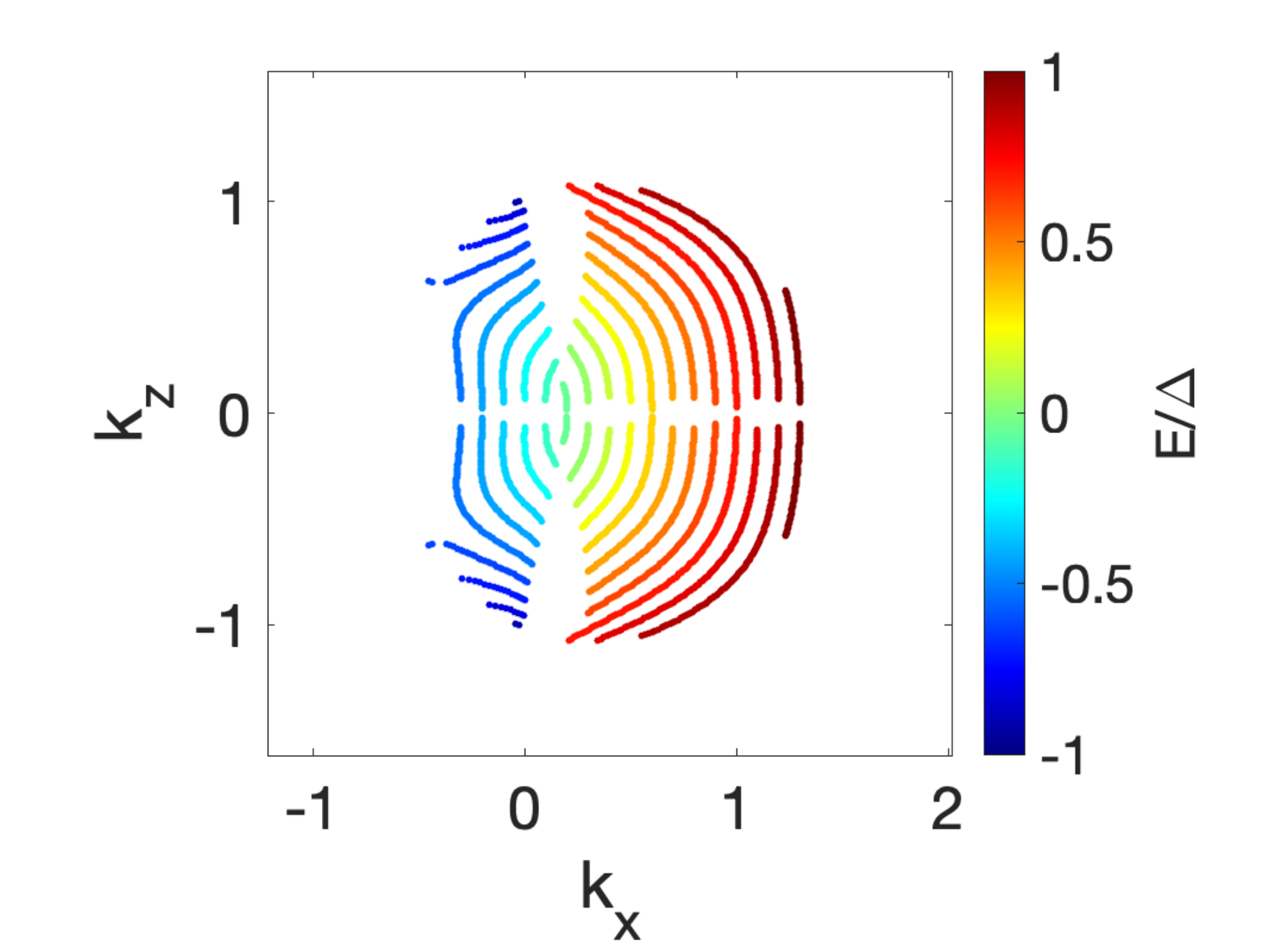}
		\caption{(Top) Arc states without SOC ($\lambda=0$, $q_{x\pm}=0$). (bottom)Arc states in the non-uniform state ($\lambda=0.8$, $q_{x+}=0.8$, $q_{x-}=-0.6$). Black dots denote the point nodes, located on $(k_x,k_z)$ plane for the chosen parameters. $m=0.5$, $\mu=1$, $\Delta_1=0.2$, $\Delta_2=\Delta_3=\Delta_4=0.15$ are used for both cases. Dashed lines are the boundary of the projected normal state Fermi surface. $\Delta=0.2$ is used for the colorbar.}
		\label{arc}
	\end{figure}

	Given a BdG Hamiltonian, the existence of chiral surface-bound states is governed by the regions in the bulk Fermi surface where the pairing gap closes. In a three-dimensional chiral p-wave state, a Fermi surface that is closed and encloses time-reversal invariant momenta will necessarily have point nodes corresponding to bulk Majorana fermion excitations.  %system with broken time-reversal symmetry (by the development of chiral p-wave state), gap closing generically happens. 
	%In the bulk Hamiltonian (without Rashba SOC), the gap closing point is the Majorana point nodes. 
	With the parameters chosen in the caption of Fig.\ref{arc}, these point nodes are located on the $(k_x,k_z)$ plane, shown as the black dots in the upper panel of Fig.\ref{arc}. They appear in pairs, with opposite momenta, due to both PH symmetry and inversion symmetry.

	For surface-bound states on the $xz$ plane, $k_x$ and $k_z$ are still good quantum numbers and label the eigenenergies, while the state decays along the y-direction into the bulk. %Eigen-energy is a function of $(k_x,k_z)$. 
	For a fixed energy $E$, the states satisfying $E(k_x,k_z)=E$ form arcs in $(k_x,k_z)$ plane. If the system has Majorana point nodes, there are zero-energy Majorana arc states, connecting two projected point nodes\cite{Kozii2016}. 
	
	\begin{figure}[h]
		\centering
		\includegraphics[width=8cm]{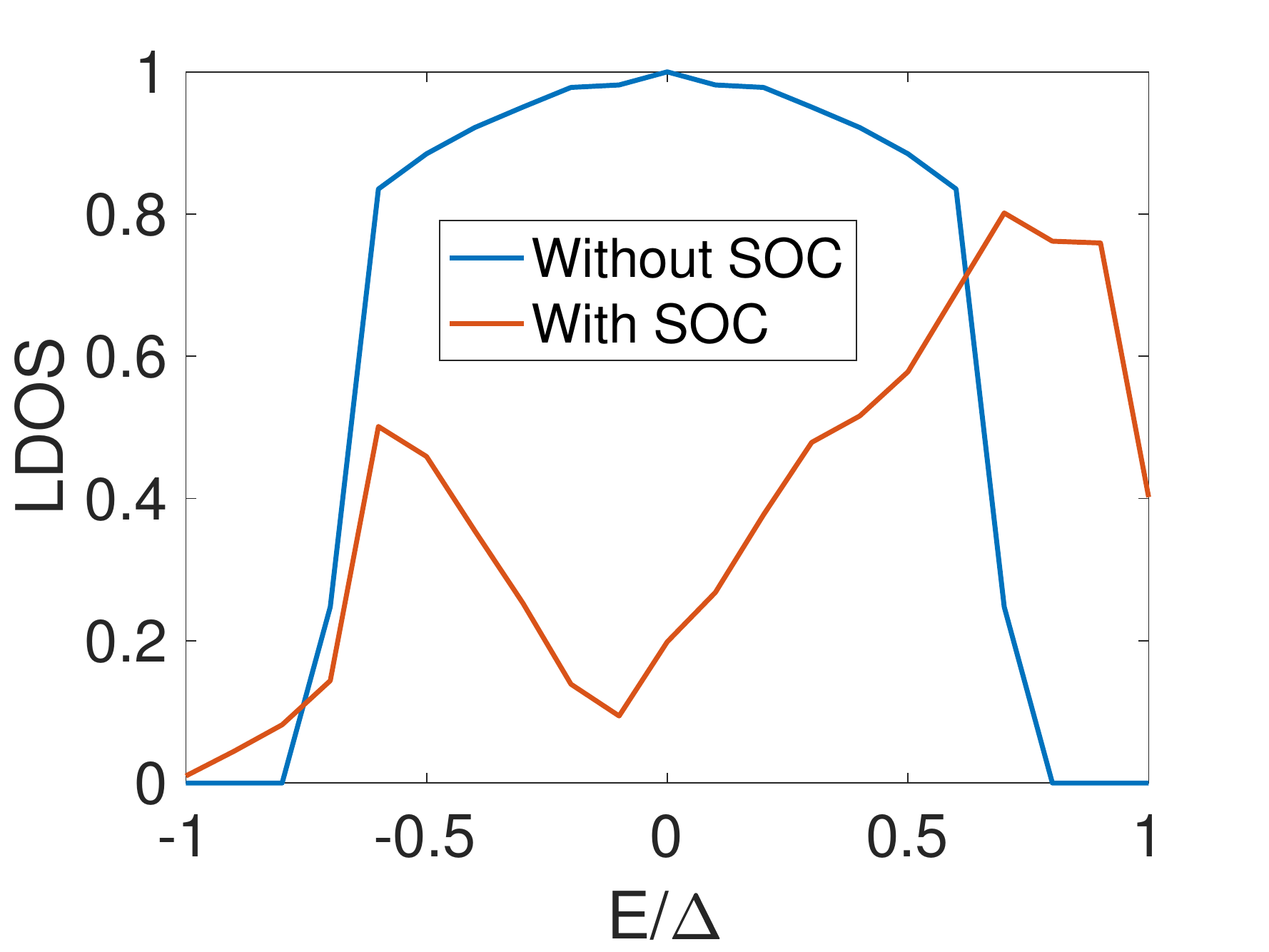}
		\caption{Density of states contributed by arc states. In the absence of Rashba SOC(blue line), system has particle-hole symmetry; and LDOS is symmetric. For non-uniform states under Rashba SOC (Red line), particle hole symmetry is broken with a peak in DOS at non-zero energy. Same parameters are used as in Fig. \ref{arc}. $\Delta\equiv0.2$ is used. }
		\label{LDOS}
	\end{figure}
	
	For the uniform $\bf q=0$ state (upper panel of Fig.\ref{arc}), zero-energy Majorana arc states (with  $E(k_x,k_z)=0$) are denoted in green lines, which are surrounded by non-zero energy arc states in other colors. Particle-hole symmetry is preserved. For example, any state on the red arc has a counterpart on the blue arc, with opposite energy and opposite momentum. For the non-uniform PDW state (lower panel), PH symmetry is broken, and there is no correspondence between positive energy states and negative energy states. 
	
	%For PDW states, since the Hamiltonian between surface and bulk is different, arc states no longer connect the physical point nodes. A natural question is: why do surface-bound states still exist? Let us start with the same assumption in the above computations, with the slowly decaying Rashba SOC, which is treated as a constant for the surface-bound states; one could consider a more general case, but the symmetry argument is the same as below. This assumption motivates us to consider a virtual bulk system with the same Hamiltonian, which will help us understand more about the surface-bound states. In this virtual bulk system, inversion symmetry is broken in the same way as near the surface; therefore it has the PDW states with the same condensate momenta. Since it is a three-dimensional system with broken time-reversal symmetry, gap closing generically happens at discrete points in the Brillouin zone. This is sufficient for the existence of surface-bound states; other symmetries (including PH symmetry) are not required. If we keep track of the surface-bound states at the middle of the gap, then we reach an arc state connecting the gap closing points with the opposite Chern number. Since gap closing generically happens at different energies, the energy of this arc has to smoothly change and connect these two energies. PH symmetry breaking is reflected in the non-zero energies of gap-closing points. If the gap closes at mostly positive energies, then surface-bound states will naturally have more positive energies than negative energies. 
	Since $k_x$ and $k_z$ are now good quantum numbers, number of states in unit square in $(k_x,k_z)$ plane is uniform. This allows us to find the quasi-particle local density of states (LDOS), contributed by arc states:
	\begin{equation} \rho_{LDOS}(E)=\sum_{k_xk_z}\delta(E-E(k_x,k_z))
	\end{equation} 
	The results can be found in Fig.\ref{LDOS}. LDOS in the uniform state is symmetric and peaked at $E=0$, while LDOS in the PDW state is asymmetric. The shape of LDOS depends on the details of the Fermi surface, the pairing function, and the orientation of the measured surface. 
	
	It should be noted that signal in STM is contributed by both surface-bound states and also bulk states, so the above analysis is far from complete. Even among the surface states, states closer to the bulk nodes will have a longer decay lengths, i.e. less localized near the surface. This may affect sensitivity to STM. Therefore, a more detailed calculation is required for the quantitative description of the STM experiment.

	\begin{figure}[h]
		\centering
		\includegraphics[width=8cm]{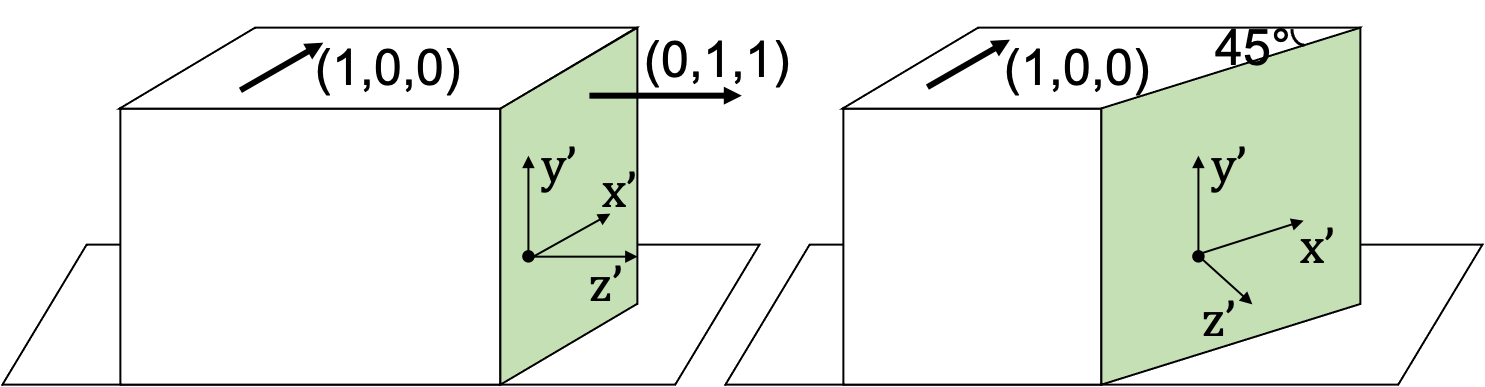}
		\caption{Experimental settings on $\text{UTe}_2$. Measurements are performed on the green surfaces. For system on the left, particle-hole symmetry is broken as it approaches the top surface. For system on the right, particle-hole symmetry is found to be unbroken. Number vectors follow the crystal axes.}
		\label{STM}
	\end{figure}

	{\it Discussions on the recent STM experiment - }%\label{S7}
	In the STM experiment of Ref. \onlinecite{jiao2020}, the LDOS near the step edge on $(0,1,1)$ and $(0,-1,-1)$ surfaces (shown in green in Fig.\ref{STM}) broke PHS. In the present context, the observations can be understood by postulating that the strength of the Rashba SOC is stronger near the step edge.  This is a reasonable hypothesis, since the confining potential is stronger, and the chiral surface modes are therefore more localized near the step edge. Thus, we expect the PDW fraction to be higher in such regions.  
	
	For LDOS on $(0,1,1)$ surface, there is a dip and peak at around opposite energies. When comparing results on $(0,1,1)$ with $(0,-1,-1)$ surface, the position of dips and peaks is reversed. Analysis on these two surfaces can be found in the appendix, but we did not find any symmetry explanations for these findings, since the only symmetry relating positive and negative energies is PH symmetry, which is broken. On the right figure of Fig.\ref{STM}, PH symmetry is found to be preserved on this ``$45^\circ$" step edge. One possible explanation is the absence of PDW since here the Rashba SOC is different from the previous surfaces. A more quantitative analysis based on realistic Fermi surface and pairing functions may explain these observations. 
 	
	%Peak positions in LDOS are opposite for $(0,1,1)$ and $(0,-1,-1)$ surfaces. Let us take the net average spin along the $(0,0,1)$ axis, as seen in the Kerr effect. This produces chiral surface modes, propagating along $\pm{}x'$ direction on these two surfaces. If these two chiral surface modes were related by $180^\circ$ rotation around $(0,1,-1)$ axis, the resulted LDOS would be the same, which will be exactly opposite to the experimental observation. Therefore, breaking this rotational symmetry is crucial. Rashba SOC from the top $(0,1,-1)$ surface breaks this rotational symmetry, while Rashba SOC from side surfaces does not. Therefore, we believe the Rashba SOC mainly comes from the top surface. However, this assumption generically leads to different, rather than exactly opposite peaks in LDOS. Another observation is the approximate PH symmetry in the $45^\circ$ measurement, on the green surface in the right panel. One possible explanation is that PDW state is not ordered in this case. Certainly, more quantitative analysis is required to fully understand these two issues.

	{\it Summary - }
	We study the effect of Rashba spin-orbit coupling on the local density of states of chiral p-wave superconductors.  We point to a natural pairing tendency towards the triplet-PDW state, in the absence of an external magnetic field and without bulk inversion symmetry breaking. We find that the LDOS near step edges shows broken PHS. Our methods are readily applicable to various experiments, including the recent STM experiment on $\text{UTe}_2$, and lend increasing support to the notion that this system hosts chiral p-wave superconductivity with Majorana fermion quasiparticle states.
	
	{\it Acknowledgments - }We thank D. Agterberg, B. Bradlyn, S.-B. Chung, S. Kivelson and F. Zhou for helpful discussions.  SR is supported  by the Department of Energy, Office of Basic Energy Sciences, Division of Materials Sciences and Engineering, under contract No. DE-AC02-76SF00515.

	\bibliography{citation}

\begin{thebibliography}{17}
\expandafter\ifx\csname natexlab\endcsname\relax\def\natexlab#1{#1}\fi
\expandafter\ifx\csname bibnamefont\endcsname\relax
  \def\bibnamefont#1{#1}\fi
\expandafter\ifx\csname bibfnamefont\endcsname\relax
  \def\bibfnamefont#1{#1}\fi
\expandafter\ifx\csname citenamefont\endcsname\relax
  \def\citenamefont#1{#1}\fi
\expandafter\ifx\csname url\endcsname\relax
  \def\url#1{\texttt{#1}}\fi
\expandafter\ifx\csname urlprefix\endcsname\relax\def\urlprefix{URL }\fi
\providecommand{\bibinfo}[2]{#2}
\providecommand{\eprint}[2][]{\url{#2}}

\bibitem[{\citenamefont{Leggett}(1975)}]{Leggett1975}
\bibinfo{author}{\bibfnamefont{A.~J.} \bibnamefont{Leggett}},
  \bibinfo{journal}{Rev. Mod. Phys.} \textbf{\bibinfo{volume}{47}},
  \bibinfo{pages}{331} (\bibinfo{year}{1975}),
  \urlprefix\url{https://link.aps.org/doi/10.1103/RevModPhys.47.331}.

\bibitem[{\citenamefont{Vollhardt and Wolfle}(2013)}]{Vollhardt2013}
\bibinfo{author}{\bibfnamefont{D.}~\bibnamefont{Vollhardt}} \bibnamefont{and}
  \bibinfo{author}{\bibfnamefont{P.}~\bibnamefont{Wolfle}},
  \emph{\bibinfo{title}{The superfluid phases of helium 3}}
  (\bibinfo{publisher}{Courier Corporation}, \bibinfo{year}{2013}).

\bibitem[{\citenamefont{Qi and Zhang}(2011)}]{Qi2011}
\bibinfo{author}{\bibfnamefont{X.-L.} \bibnamefont{Qi}} \bibnamefont{and}
  \bibinfo{author}{\bibfnamefont{S.-C.} \bibnamefont{Zhang}},
  \bibinfo{journal}{Reviews of Modern Physics} \textbf{\bibinfo{volume}{83}},
  \bibinfo{pages}{1057} (\bibinfo{year}{2011}).

\bibitem[{\citenamefont{Leijnse and Flensberg}(2012)}]{leijnse2012}
\bibinfo{author}{\bibfnamefont{M.}~\bibnamefont{Leijnse}} \bibnamefont{and}
  \bibinfo{author}{\bibfnamefont{K.}~\bibnamefont{Flensberg}},
  \bibinfo{journal}{Semiconductor Science and Technology}
  \textbf{\bibinfo{volume}{27}}, \bibinfo{pages}{124003}
  (\bibinfo{year}{2012}).

\bibitem[{\citenamefont{Sato and Ando}(2017)}]{Sato2017}
\bibinfo{author}{\bibfnamefont{M.}~\bibnamefont{Sato}} \bibnamefont{and}
  \bibinfo{author}{\bibfnamefont{Y.}~\bibnamefont{Ando}},
  \bibinfo{journal}{Reports on Progress in Physics}
  \textbf{\bibinfo{volume}{80}}, \bibinfo{pages}{076501}
  (\bibinfo{year}{2017}),
  \urlprefix\url{https://doi.org/10.1088/1361-6633/aa6ac7}.

\bibitem[{\citenamefont{Cheung and Raghu}(2016)}]{Cheung2016}
\bibinfo{author}{\bibfnamefont{A.~K.~C.} \bibnamefont{Cheung}}
  \bibnamefont{and} \bibinfo{author}{\bibfnamefont{S.}~\bibnamefont{Raghu}},
  \bibinfo{journal}{Phys. Rev. B} \textbf{\bibinfo{volume}{93}},
  \bibinfo{pages}{134516} (\bibinfo{year}{2016}),
  \urlprefix\url{https://link.aps.org/doi/10.1103/PhysRevB.93.134516}.

\bibitem[{\citenamefont{Sigrist and Ueda}(1991)}]{Sigrist1991}
\bibinfo{author}{\bibfnamefont{M.}~\bibnamefont{Sigrist}} \bibnamefont{and}
  \bibinfo{author}{\bibfnamefont{K.}~\bibnamefont{Ueda}},
  \bibinfo{journal}{Rev. Mod. Phys.} \textbf{\bibinfo{volume}{63}},
  \bibinfo{pages}{239} (\bibinfo{year}{1991}),
  \urlprefix\url{https://link.aps.org/doi/10.1103/RevModPhys.63.239}.

\bibitem[{\citenamefont{Aoki et~al.}(2019)\citenamefont{Aoki, Nakamura, Honda,
  Li, Homma, Shimizu, Sato, Knebel, Brison, Pourret et~al.}}]{aoki2019}
\bibinfo{author}{\bibfnamefont{D.}~\bibnamefont{Aoki}},
  \bibinfo{author}{\bibfnamefont{A.}~\bibnamefont{Nakamura}},
  \bibinfo{author}{\bibfnamefont{F.}~\bibnamefont{Honda}},
  \bibinfo{author}{\bibfnamefont{D.}~\bibnamefont{Li}},
  \bibinfo{author}{\bibfnamefont{Y.}~\bibnamefont{Homma}},
  \bibinfo{author}{\bibfnamefont{Y.}~\bibnamefont{Shimizu}},
  \bibinfo{author}{\bibfnamefont{Y.~J.} \bibnamefont{Sato}},
  \bibinfo{author}{\bibfnamefont{G.}~\bibnamefont{Knebel}},
  \bibinfo{author}{\bibfnamefont{J.-P.} \bibnamefont{Brison}},
  \bibinfo{author}{\bibfnamefont{A.}~\bibnamefont{Pourret}},
  \bibnamefont{et~al.}, \bibinfo{journal}{Journal of the Physical Society of
  Japan} \textbf{\bibinfo{volume}{88}}, \bibinfo{pages}{043702}
  (\bibinfo{year}{2019}).

\bibitem[{\citenamefont{Ran et~al.}(2019)\citenamefont{Ran, Liu, Eo, Campbell,
  Neves, Fuhrman, Saha, Eckberg, Kim, Graf et~al.}}]{ran2019}
\bibinfo{author}{\bibfnamefont{S.}~\bibnamefont{Ran}},
  \bibinfo{author}{\bibfnamefont{I.-L.} \bibnamefont{Liu}},
  \bibinfo{author}{\bibfnamefont{Y.~S.} \bibnamefont{Eo}},
  \bibinfo{author}{\bibfnamefont{D.~J.} \bibnamefont{Campbell}},
  \bibinfo{author}{\bibfnamefont{P.~M.} \bibnamefont{Neves}},
  \bibinfo{author}{\bibfnamefont{W.~T.} \bibnamefont{Fuhrman}},
  \bibinfo{author}{\bibfnamefont{S.~R.} \bibnamefont{Saha}},
  \bibinfo{author}{\bibfnamefont{C.}~\bibnamefont{Eckberg}},
  \bibinfo{author}{\bibfnamefont{H.}~\bibnamefont{Kim}},
  \bibinfo{author}{\bibfnamefont{D.}~\bibnamefont{Graf}}, \bibnamefont{et~al.},
  \bibinfo{journal}{Nature Physics} \textbf{\bibinfo{volume}{15}},
  \bibinfo{pages}{1250} (\bibinfo{year}{2019}).

\bibitem[{\citenamefont{Hayes et~al.}(2020)\citenamefont{Hayes, Wei, Metz,
  Zhang, Eo, Ran, Saha, Collini, Butch, Agterberg et~al.}}]{hayes2020}
\bibinfo{author}{\bibfnamefont{I.~M.} \bibnamefont{Hayes}},
  \bibinfo{author}{\bibfnamefont{D.~S.} \bibnamefont{Wei}},
  \bibinfo{author}{\bibfnamefont{T.}~\bibnamefont{Metz}},
  \bibinfo{author}{\bibfnamefont{J.}~\bibnamefont{Zhang}},
  \bibinfo{author}{\bibfnamefont{Y.~S.} \bibnamefont{Eo}},
  \bibinfo{author}{\bibfnamefont{S.}~\bibnamefont{Ran}},
  \bibinfo{author}{\bibfnamefont{S.~R.} \bibnamefont{Saha}},
  \bibinfo{author}{\bibfnamefont{J.}~\bibnamefont{Collini}},
  \bibinfo{author}{\bibfnamefont{N.~P.} \bibnamefont{Butch}},
  \bibinfo{author}{\bibfnamefont{D.~F.} \bibnamefont{Agterberg}},
  \bibnamefont{et~al.}, \bibinfo{journal}{arXiv preprint arXiv:2002.02539}
  (\bibinfo{year}{2020}).

\bibitem[{\citenamefont{Jiao et~al.}(2020)\citenamefont{Jiao, Howard, Ran,
  Wang, Rodriguez, Sigrist, Wang, Butch, and Madhavan}}]{jiao2020}
\bibinfo{author}{\bibfnamefont{L.}~\bibnamefont{Jiao}},
  \bibinfo{author}{\bibfnamefont{S.}~\bibnamefont{Howard}},
  \bibinfo{author}{\bibfnamefont{S.}~\bibnamefont{Ran}},
  \bibinfo{author}{\bibfnamefont{Z.}~\bibnamefont{Wang}},
  \bibinfo{author}{\bibfnamefont{J.~O.} \bibnamefont{Rodriguez}},
  \bibinfo{author}{\bibfnamefont{M.}~\bibnamefont{Sigrist}},
  \bibinfo{author}{\bibfnamefont{Z.}~\bibnamefont{Wang}},
  \bibinfo{author}{\bibfnamefont{N.~P.} \bibnamefont{Butch}}, \bibnamefont{and}
  \bibinfo{author}{\bibfnamefont{V.}~\bibnamefont{Madhavan}},
  \bibinfo{journal}{Nature} \textbf{\bibinfo{volume}{579}},
  \bibinfo{pages}{523} (\bibinfo{year}{2020}).

\bibitem[{\citenamefont{Fulde and Ferrell}(1964)}]{FF1964}
\bibinfo{author}{\bibfnamefont{P.}~\bibnamefont{Fulde}} \bibnamefont{and}
  \bibinfo{author}{\bibfnamefont{R.~A.} \bibnamefont{Ferrell}},
  \bibinfo{journal}{Phys. Rev.} \textbf{\bibinfo{volume}{135}},
  \bibinfo{pages}{A550} (\bibinfo{year}{1964}),
  \urlprefix\url{https://link.aps.org/doi/10.1103/PhysRev.135.A550}.

\bibitem[{\citenamefont{Larkin and Ovchinnikov}(1965)}]{LO1965}
\bibinfo{author}{\bibfnamefont{A.}~\bibnamefont{Larkin}} \bibnamefont{and}
  \bibinfo{author}{\bibfnamefont{I.}~\bibnamefont{Ovchinnikov}},
  \bibinfo{journal}{Soviet Physics-JETP} \textbf{\bibinfo{volume}{20}},
  \bibinfo{pages}{762} (\bibinfo{year}{1965}).

\bibitem[{\citenamefont{Agterberg et~al.}(2020)\citenamefont{Agterberg, Davis,
  Edkins, Fradkin, Van~Harlingen, Kivelson, Lee, Radzihovsky, Tranquada, and
  Wang}}]{Agterberg2020}
\bibinfo{author}{\bibfnamefont{D.~F.} \bibnamefont{Agterberg}},
  \bibinfo{author}{\bibfnamefont{J.~S.} \bibnamefont{Davis}},
  \bibinfo{author}{\bibfnamefont{S.~D.} \bibnamefont{Edkins}},
  \bibinfo{author}{\bibfnamefont{E.}~\bibnamefont{Fradkin}},
  \bibinfo{author}{\bibfnamefont{D.~J.} \bibnamefont{Van~Harlingen}},
  \bibinfo{author}{\bibfnamefont{S.~A.} \bibnamefont{Kivelson}},
  \bibinfo{author}{\bibfnamefont{P.~A.} \bibnamefont{Lee}},
  \bibinfo{author}{\bibfnamefont{L.}~\bibnamefont{Radzihovsky}},
  \bibinfo{author}{\bibfnamefont{J.~M.} \bibnamefont{Tranquada}},
  \bibnamefont{and} \bibinfo{author}{\bibfnamefont{Y.}~\bibnamefont{Wang}},
  \bibinfo{journal}{Annual Review of Condensed Matter Physics}
  \textbf{\bibinfo{volume}{11}}, \bibinfo{pages}{231–270}
  (\bibinfo{year}{2020}), ISSN \bibinfo{issn}{1947-5462},
  \urlprefix\url{http://dx.doi.org/10.1146/annurev-conmatphys-031119-050711}.

\bibitem[{\citenamefont{Martin et~al.}(2005)\citenamefont{Martin, Agosta,
  Tozer, Radovan, Palm, Murphy, and Sarrao}}]{martin2005}
\bibinfo{author}{\bibfnamefont{C.}~\bibnamefont{Martin}},
  \bibinfo{author}{\bibfnamefont{C.}~\bibnamefont{Agosta}},
  \bibinfo{author}{\bibfnamefont{S.}~\bibnamefont{Tozer}},
  \bibinfo{author}{\bibfnamefont{H.}~\bibnamefont{Radovan}},
  \bibinfo{author}{\bibfnamefont{E.}~\bibnamefont{Palm}},
  \bibinfo{author}{\bibfnamefont{T.}~\bibnamefont{Murphy}}, \bibnamefont{and}
  \bibinfo{author}{\bibfnamefont{J.}~\bibnamefont{Sarrao}},
  \bibinfo{journal}{Physical Review B} \textbf{\bibinfo{volume}{71}},
  \bibinfo{pages}{020503} (\bibinfo{year}{2005}).

\bibitem[{\citenamefont{Kenzelmann et~al.}(2008)\citenamefont{Kenzelmann,
  Str{\"a}ssle, Niedermayer, Sigrist, Padmanabhan, Zolliker, Bianchi,
  Movshovich, Bauer, Sarrao et~al.}}]{Kenzelmann2008}
\bibinfo{author}{\bibfnamefont{M.}~\bibnamefont{Kenzelmann}},
  \bibinfo{author}{\bibfnamefont{T.}~\bibnamefont{Str{\"a}ssle}},
  \bibinfo{author}{\bibfnamefont{C.}~\bibnamefont{Niedermayer}},
  \bibinfo{author}{\bibfnamefont{M.}~\bibnamefont{Sigrist}},
  \bibinfo{author}{\bibfnamefont{B.}~\bibnamefont{Padmanabhan}},
  \bibinfo{author}{\bibfnamefont{M.}~\bibnamefont{Zolliker}},
  \bibinfo{author}{\bibfnamefont{A.~D.} \bibnamefont{Bianchi}},
  \bibinfo{author}{\bibfnamefont{R.}~\bibnamefont{Movshovich}},
  \bibinfo{author}{\bibfnamefont{E.~D.} \bibnamefont{Bauer}},
  \bibinfo{author}{\bibfnamefont{J.~L.} \bibnamefont{Sarrao}},
  \bibnamefont{et~al.}, \bibinfo{journal}{Science}
  \textbf{\bibinfo{volume}{321}}, \bibinfo{pages}{1652} (\bibinfo{year}{2008}),
  ISSN \bibinfo{issn}{0036-8075},
  \eprint{https://science.sciencemag.org/content/321/5896/1652.full.pdf},
  \urlprefix\url{https://science.sciencemag.org/content/321/5896/1652}.

\bibitem[{\citenamefont{Kozii et~al.}(2016)\citenamefont{Kozii, Venderbos, and
  Fu}}]{Kozii2016}
\bibinfo{author}{\bibfnamefont{V.}~\bibnamefont{Kozii}},
  \bibinfo{author}{\bibfnamefont{J.~W.~F.} \bibnamefont{Venderbos}},
  \bibnamefont{and} \bibinfo{author}{\bibfnamefont{L.}~\bibnamefont{Fu}},
  \bibinfo{journal}{Science Advances} \textbf{\bibinfo{volume}{2}}
  (\bibinfo{year}{2016}),
  \eprint{https://advances.sciencemag.org/content/2/12/e1601835.full.pdf},
  \urlprefix\url{https://advances.sciencemag.org/content/2/12/e1601835}.

\end{thebibliography}

	\onecolumngrid
	\clearpage
	\appendix
	\renewcommand\thefigure{\thesection.\arabic{figure}}

	\section{perfect nesting for spherical Fermi surface}
\begin{figure}[h]
		\centering
		\includegraphics[width=7cm]{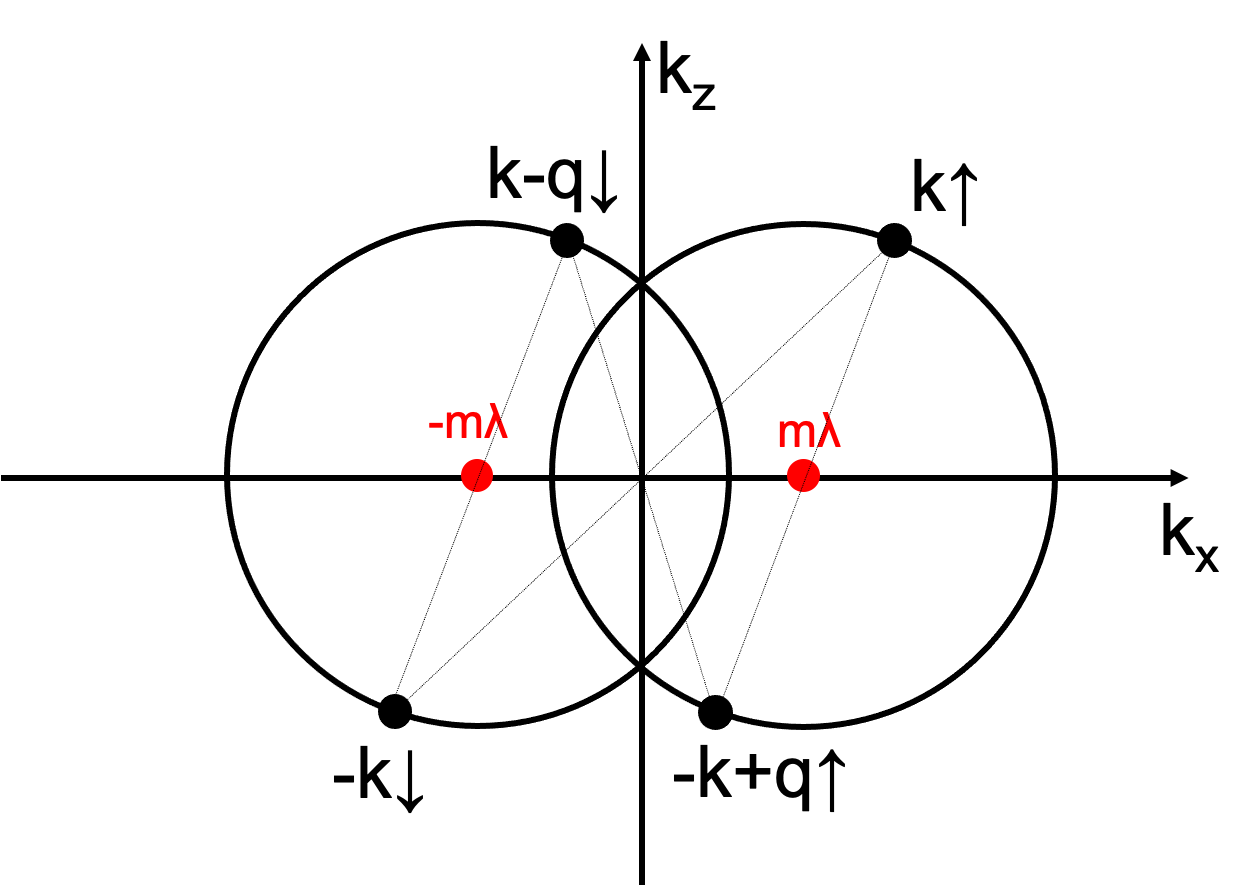}
		\caption{Spherical Fermi surface with $\lambda'=0$, projected to $(k_x,k_z)$ plane. Perfect nesting with $q_{x+}=-q_{x-}=2m\lambda$ is shown by the dotted line.}
		\label{FS}
	\end{figure}

	A special example for PDW state is the elliptical Fermi surface with $\varepsilon_{\bf k}=\sum_i\frac{k_i^2}{2m_i}$. The spin-up and spin-down Fermi surfaces are shifted by $\pm{m_x\lambda}$, as shown in Fig.\ref{FS}. Due to the quadratic dispersion, the final Fermi surfaces are not distorted, which leads to a new Fermi surface perfect nesting. Pairing susceptibility diverges at $q_{x+}=-q_{x-}=2m_x\lambda$ and $q_{x0}=0$. In the calculations for surface-bound states, we used a spherical Fermi surface. But we did not focus on Fermi surface perfect nesting, i.e. $\bf q_{\pm,0}$ are taken to be independent parameters in the model.

	\begin{figure}[h]
		\centering
		\includegraphics[width=7cm]{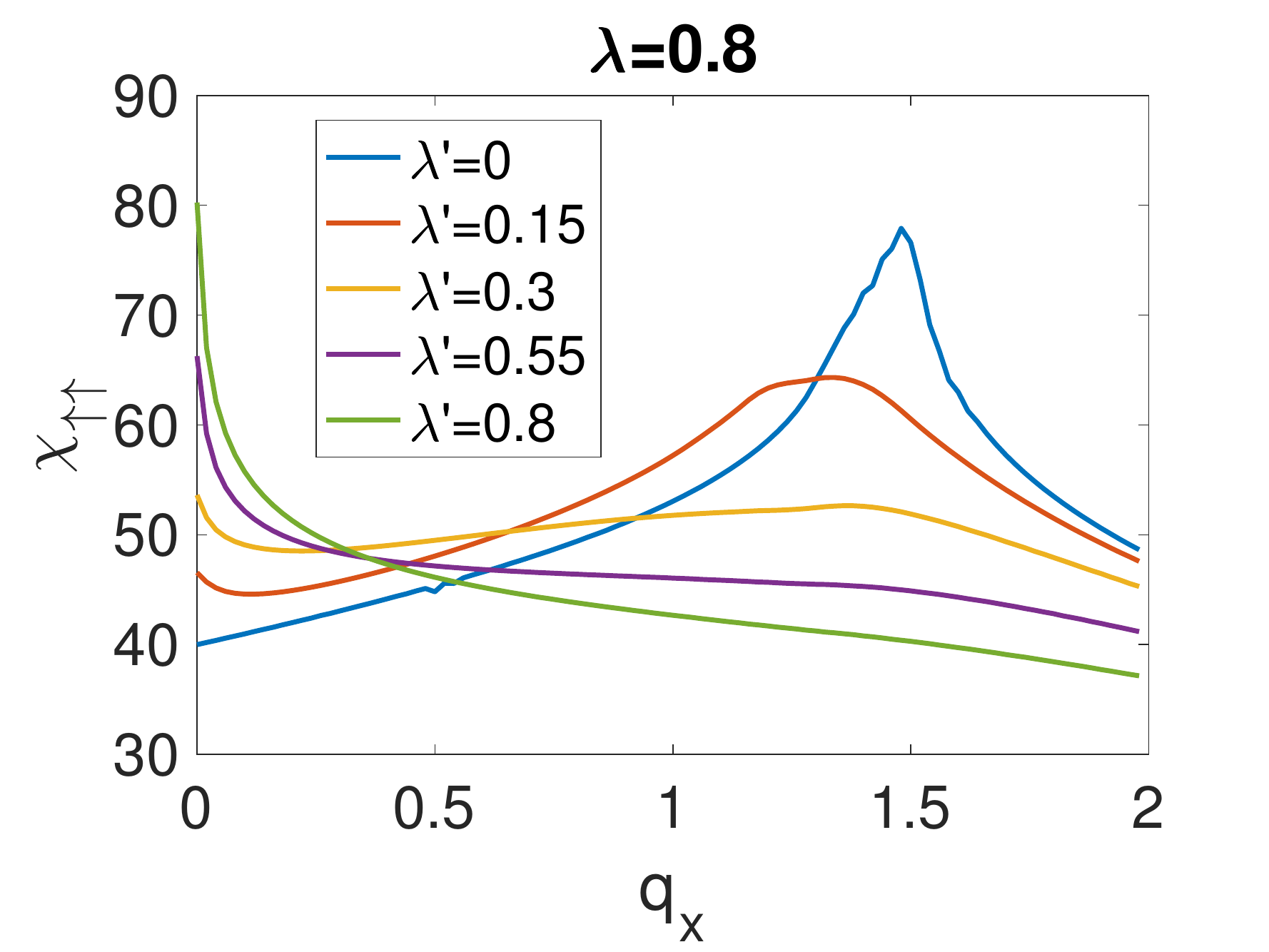}
		\caption{Pairing susceptibility for fixed $\lambda$ with various $\lambda'$}
		\label{chi}
	\end{figure}
	\section{pairing susceptibility}
	For simplicity, we calculated pairing susceptibilities for state $\Delta_{\uparrow\uparrow}^{\bf q}({\bf k})=k_z$, with Cooper pair momentum ${\bf q}=(q_{x},0,0)$, while neglecting all other states. The same Hamiltonian and parameters are used as in the caption of Fig.\ref{FS0}. The results are shown in Fig.\ref{chi}. We found a critical $\lambda'$ around 0.3. For $\lambda'<0.3$, pairing susceptibility is maximized at non-zero momentum. For $\lambda'>0.3$, it is maximized at zero momentum. It is worth noting that, other states certainly play a more important role for larger $\lambda'$, and more quantitative analysis including pairing interactions is needed to get a complete phase diagram.

	\section{Translational symmetry\&coupling to charge-density-wave}
	Condensate with a single center of mass momentum preserves particle-hole symmetry, as one can always shift the origin of momentum space and let $\bf q=0$. To break the translational symmetry and particle-hole symmetry, it is necessary to have at least two condensate momenta. The difference in these momenta determines the orientation and periodicity of PDW:
	\begin{equation} \Delta(x)\sim\cos(\frac{1}{2}\Delta\bf{q\cdot{x}})
	\end{equation}
	The model generically has more than two condensate momenta. For instance, in the limit of $\lambda'\ll\lambda$, PDW has one Cooper pair momentum in each pairing channel (if ordered). Therefore, we will require at least two pairing channels to be ordered for PH symmetry breaking. An easier possibility is to have ordered $\Delta_{\uparrow\uparrow}$ and $\Delta_{\downarrow\downarrow}$ channels since the competition between them is weak. Other coexisting scenarios will require stronger pairing interactions.
	
	Next, we discuss a consequence of non-uniform superconductivity: the development of charge modulation. The coupling between the charge density wave (CDW) and superconducting order paramaters can be deduced from Ginzburg-Landau theory.  Gauge invariance constrains the lowest order coupling to be quadratic in $\Delta$:
	\begin{equation}
		\begin{split}
			f_{int}=\sum_{ijab}\alpha^{ij}_{ab}\rho_{\bf q_j-q_i}\left[\Delta_a^{\bf q_i}(\Delta_b^{\bf q_j})^*+(\Delta_{a'}^{\bf -q_i})^*\Delta_{b'}^{\bf -q_j}\right]
		\end{split}
	\end{equation}
	, with proper coefficients $\alpha_{ab}^{ij}$ determined by the normal state dispersion. $a,b$ denotes different pairing channels: $a,b=\uparrow\uparrow,\uparrow\downarrow,\downarrow\downarrow$, while their time-reversal pairs are $a',b'=\downarrow\downarrow,\uparrow\downarrow,\uparrow\uparrow$. 
	
	\subsection{special case at $\lambda'=0$}
	But right at $\lambda'=0$, translational symmetry is more subtle. Since the normal state Hamiltonian does not couple spin-up and spin-down states, different channels do not couple in the above equation; i.e. $\alpha_{ab}^{ij}\neq0$ only for $a=b$. Since there is a unique $\bf q$ in each pairing channel, the above quadratic terms only couple to uniform charge density, rather than CDW. The actual leading order contribution to CDW is then quartic in $\Delta$:
	\begin{equation}
		\begin{split}
			&\rho_{\bf q}^{\uparrow\uparrow},\rho_{\bf q}^{\downarrow\downarrow}\propto\left[\Delta^{\uparrow\uparrow}_{\bf q_+}\Delta^{\uparrow\downarrow*}_{\bf q_0}\Delta^{\downarrow\downarrow}_{\bf q_-}\Delta^{\uparrow\downarrow*}_{\bf q_0}+\Delta^{\uparrow\uparrow*}_{\bf -q_+}\Delta^{\uparrow\downarrow}_{\bf -q_0}\Delta^{\downarrow\downarrow*}_{\bf -q_-}\Delta^{\uparrow\downarrow}_{\bf -q_0}\right]\\
			&{\bf q=q_++q_--2q_0},
		\end{split}
	\end{equation}
	A non-zero wavevector $\bf q$ requires $\bf q_++q_-\neq2q_0$, which is the condition of translational symmetry breaking at $\lambda'=0$. CDW is expected to be weak at $\lambda'=0$ since it is only contributed by quartic terms in $\Delta$. CDW should be still small in the limit of $\lambda'<<\lambda$.

	\section{$(0,-1,-1)$ and $(0,1,1)$ surfaces}
	\setcounter{figure}{0} 
	In this section, $(x,y,z)$ axes will denote the crystal axes, while $(x',y',z')$ axes are local coordinate set up in Fig.\ref{STM}. The pairing state is taken to be $k_x+ik_y$ state, satisfying the Kerr effect measurement, with the following gap function 
	\begin{equation}
		\begin{split}
			&\vec{d}_0=(\Delta_3k_z,i\Delta_4k_z,\Delta_1k_x+i\Delta_2k_y)
		\end{split}
	\end{equation}
	
	As we consider surface bound state on $(0,-1,-1)$ and $(0,1,1)$ surfaces, we need to go to the local basis, where y'-axis is normal vector of the top surface, while z' is the normal vector of the side $(0,1,1)$/$(0,-1,-1)$ surface. x'-axis is the same as crystal x-axis. Basis transformation leads to a vector rotation on both $\bf k$ and a vector rotation of the d-vector:
	\begin{equation}
		\begin{split}
			&\vec{d}({\bf k'})=M\vec{d}_0(M{\bf k})
		\end{split}
	\end{equation}
	, where matrix $M$ is the standard rotation matrix in 3D.
	
	Now we can calculate surface-bound states on the two side surfaces, and the results is shown below:
	
	\begin{figure}[h]
		\centering
		\includegraphics[width=8cm]{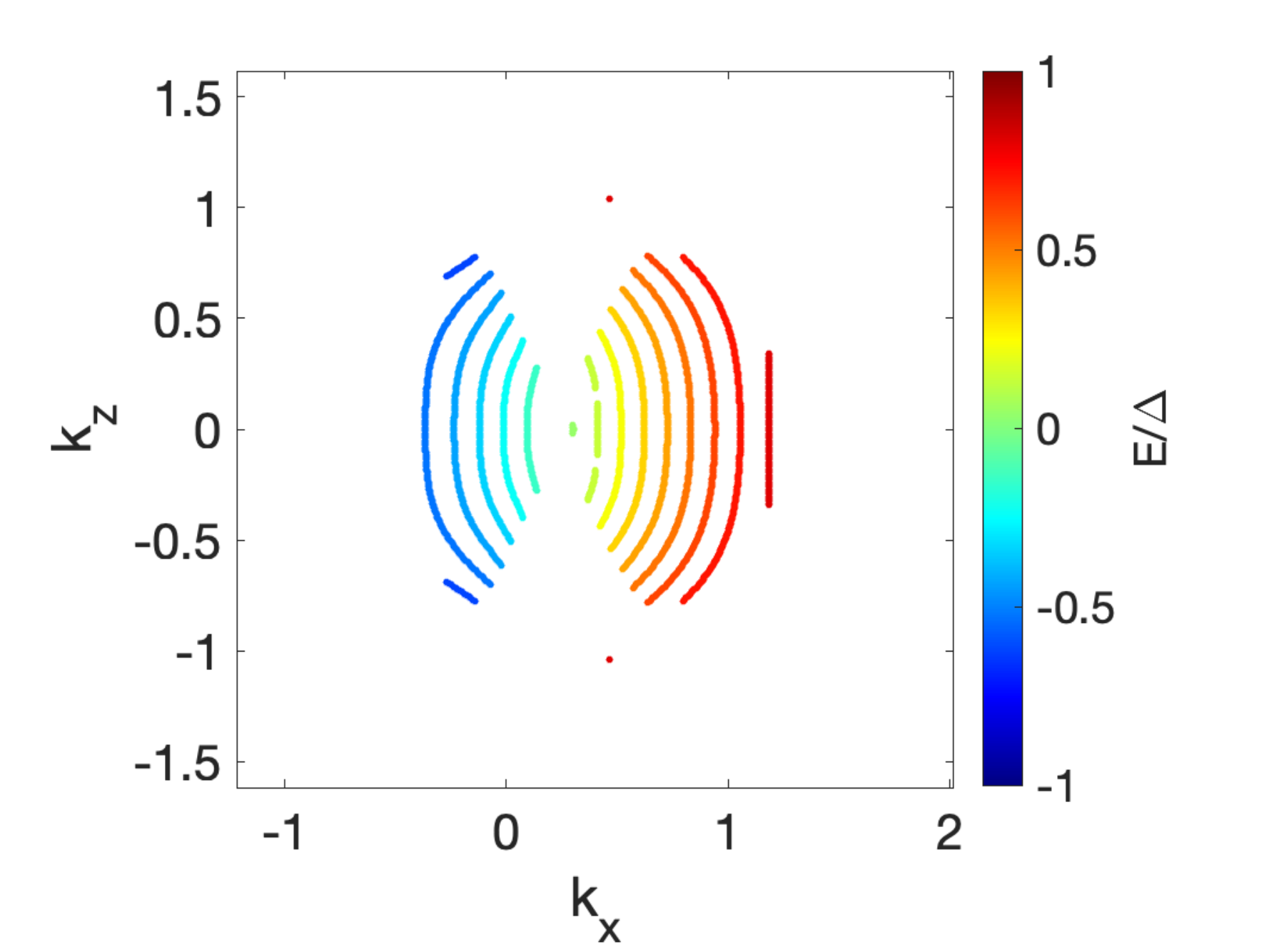}
		\includegraphics[width=8cm]{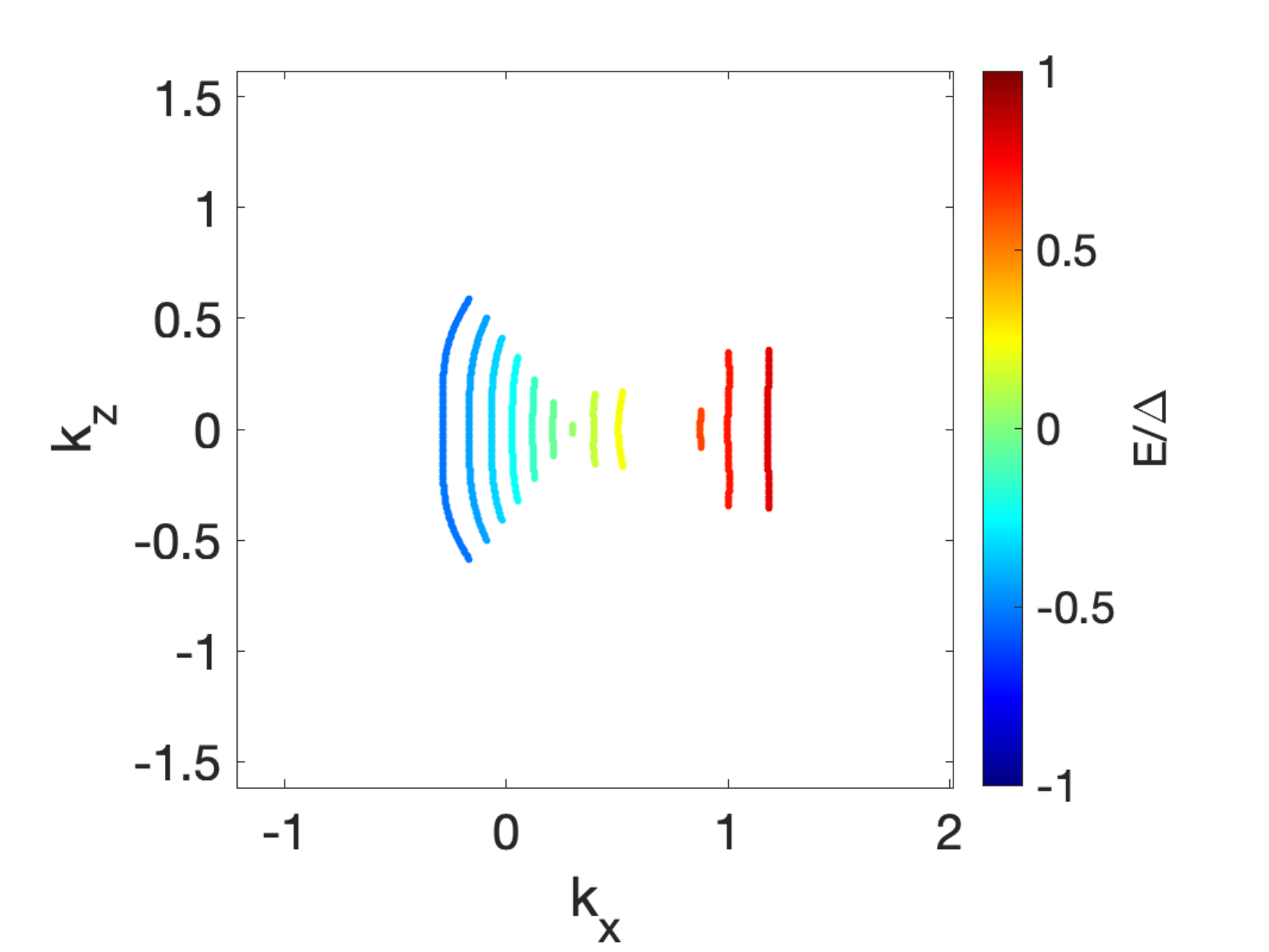}
		\caption{(left) (0,1,1) surface. Same parameters $\Delta_i$ for pairing states were used as in main text.}
	\end{figure}
	We observed that the peak in LDOS on the two surfaces should be located at different energies. But LDOS profile is not the same.

	\section{Truncation to larger matrix}
	Here we would like to show the surface bound states with truncation to $8\times8$ Hamiltonian:
	\begin{equation}
		\begin{split}
			&H=\sum_{\bf k}\Phi_{\bf k}^\dagger{}h_{\bf k}\Phi_{\bf k}\\
			&\Phi_{\bf k}^\dagger=[c_{\bf k\uparrow}^\dagger\;\; 
			c_{\bf k-q_+\downarrow}^\dagger\;\;
			c_{\bf k+q_-\downarrow}^\dagger\;\;
			c_{\bf k+q_++q_-\uparrow}^\dagger\;\;
			c_{\bf -k+q_+\uparrow}\;\;
			c_{\bf -k\downarrow}\;\;
			c_{\bf -k+q_++q_-\downarrow}\;\;
			c_{\bf -k-q_-\uparrow}]\\
			&h_{\bf k}=\left[
			\begin{array}{cccccccc}
				\varepsilon_{\bf k\uparrow} &&&& \Delta_{\uparrow\uparrow}({\bf k}) & \Delta_{\uparrow\downarrow}({\bf k})&&\\
				&\varepsilon_{\bf k-q_+\downarrow}
				&&&\Delta_{\uparrow\downarrow}(\bf k-q_+)
				&&\Delta_{\downarrow\downarrow}(\bf k-q_+)& \\
				&& \varepsilon_{\bf k+q_-\downarrow} &&& \Delta_{\downarrow\downarrow}({\bf k+q_-})&&
				\Delta_{\uparrow\downarrow}({\bf k+q_-})\\
				&&& \varepsilon_{\bf k+q_++q_-\uparrow} &&&&
				\Delta_{\uparrow\downarrow}({\bf k+q_++q_-})\\
				\Delta^*_{\uparrow\uparrow}({\bf k}) &
				\Delta^*_{\uparrow\downarrow}({\bf k-q_+})&&& -\varepsilon_{\bf -k+q_+\uparrow} \\
				\Delta^*_{\uparrow\downarrow}({\bf k}) &&
				\Delta^*_{\downarrow\downarrow}({\bf k+q_-}) &&& -\varepsilon_{\bf -k\downarrow}\\
				&\Delta^*_{\downarrow\downarrow}({\bf k-q_+})&&&&&
				-\varepsilon_{\bf -k+q_++q_-\downarrow}\\
				&&\Delta^*_{\uparrow\downarrow}({\bf k+q_-})&
				\Delta^*_{\uparrow\downarrow}({\bf k+q_++q_-})&&&&
				-\varepsilon_{\bf -k-q_-\uparrow}
			\end{array}
			\right]
		\end{split}
	\end{equation}
	
	The result is shown below. 
	\begin{figure}[h]
		\centering
		\includegraphics[width=8cm]{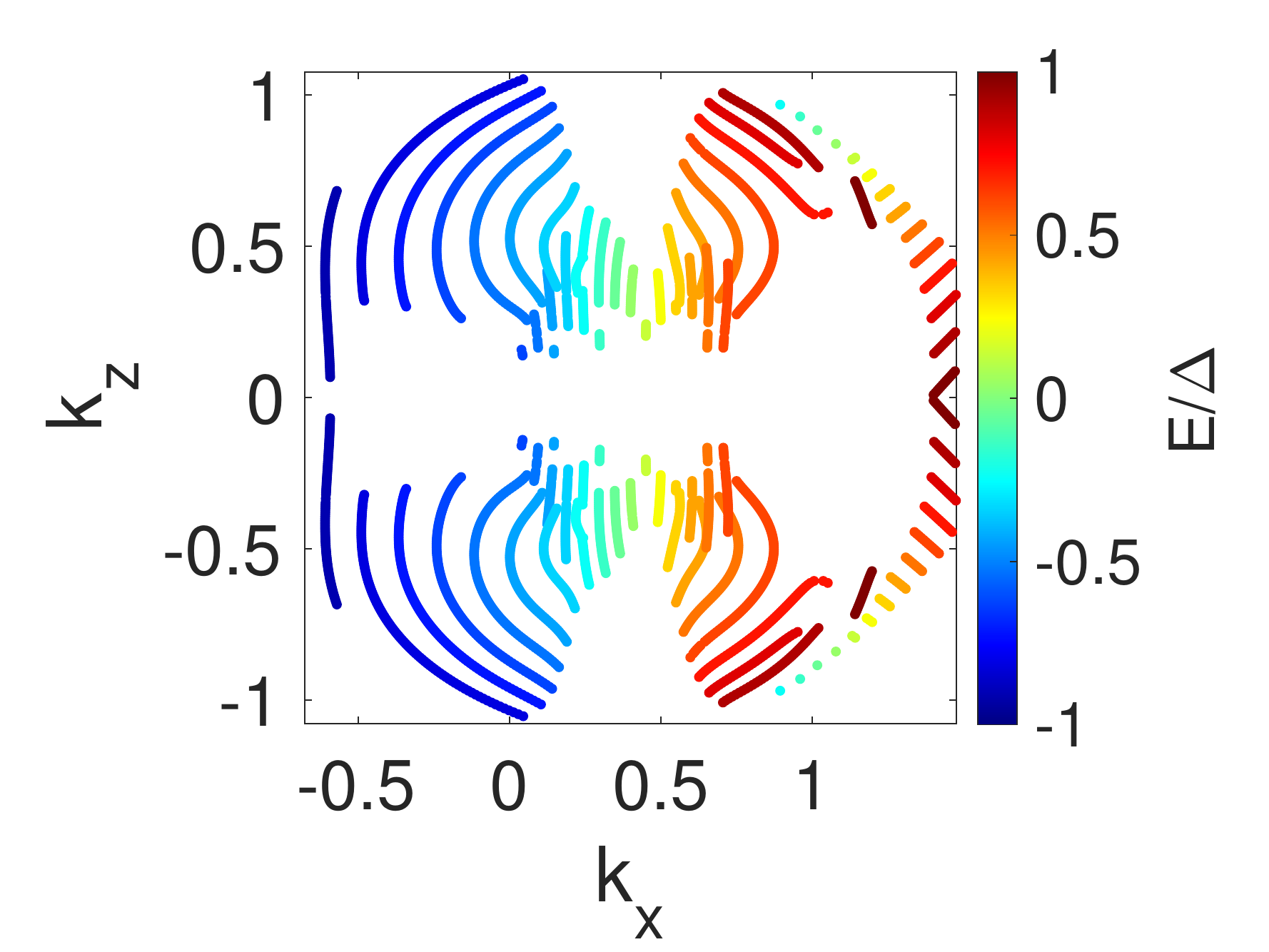}
		\caption{surface bound states of $8\times8$ truncation. Same parameters are used as in main text. }
	\end{figure}

\end{document}